\documentclass[aps,reprint,pre,showpacs,floatfix]{revtex4-2}
\usepackage{times,amsmath,graphicx,latexsym}

\begin{document}

\title{Coulomb gas sum rules for vortex-pair fluctuations in 2D superfluids
}

\author{Mingyu Fan}
\affiliation{Department of Physics and Astronomy, University of California,
Los Angeles, CA 90095}
\author{Karla Galdamez} 
\affiliation{Department of Chemistry and Biochemistry, University of California,
Santa Cruz, CA 95064} 
\author{Charlie McDowell} 
\affiliation{Department of Computer Science and Engineering, University of California,
Santa Cruz, CA 95064}
\author{Gary A. Williams}
\email{gaw@ucla.edu}
\affiliation{Department of Physics and Astronomy, University of California,
Los Angeles, CA 90095}
\date{\today}
\begin{abstract} 
Vortex fluctuations above and below the critical Kosterlitz-Thouless (KT) transition temperature are characterized using simulations of the 2D XY model.  The net winding number of vortices at a given temperature in a circle of radius $R$ is computed as a function of $R$.   The average squared winding number is found to vary linearly with the perimeter of the circle at all temperatures above and below $T_{KT}$, and the slope with $R$ displays a sharp peak near the specific heat peak, decreasing then to a value at infinite temperature that is in agreement with an early theory by Dhar.   We have also computed the vortex-vortex distribution functions, finding an asymptotic power-law variation in the vortex separation distance at all temperatures.  In conjunction with a Coulomb-gas sum rule on the perimeter fluctuations, these can be used to successfully model the start of the perimeter-slope peak in the region below $T_{KT}$.

 \end{abstract}  

\maketitle

The role of topological vortices in phase transitions in two-dimensional (2D) superfluids was first explored by  Kosterlitz and Thouless (KT) \cite{kt,kosterlitzrev} and Berezhinskii \cite{berez}.  A real-space renormalization  group theory by Kosterlitz \cite{kosterlitz} was able to characterize many of the properties of the phase transition below the critical temperature $T_{KT}$, and simulations of 2D XY spin lattices were able to verify many aspects of the theory \cite{tobochnik,hasen}.  However, the recursion relations of the theory blow up at temperatures above $T_{KT}$, making it difficult to explore what happens in this region where the vortex density becomes large, approaching an average density of 1/3 per lattice site as $T \rightarrow \infty$.   Kosterlitz was able to calculate the correlation length above the transition, an exponential decrease from the infinite value below $T_{KT}$,
\begin{equation}
\xi  = {\xi _0}\,{e^{b\,{((T - T_{KT})/T_{KT})^{ - 1/2}}}}
\end{equation}
where $b$ and $\xi_0$ are nonuniversal constants; XY simulations gave $b$ = 1.53 and $\xi_0$ on the order of the core size \cite{binder}.  This is often interpreted as characterizing the ``unbinding" of the vortex pairs with increasing temperature, giving the vortex-antivortex separation of the pairs still bound above $T_{KT}$.   The specific heat above the transition was also calculated \cite{berker} by extending the renormalization theory using a Debye-Huckle approximation (Eq.\,1 can also be interpreted as a Debye screening length), and a peak in the specific heat was found at $T/T_{KT}$ = 1.3 from the unbinding of the vortex pairs of smallest separation.  Simulations showed a similar peak \cite{tobochnik,pre_sh}, but actually occurring at a lower peak temperature of about 1.18 $T_{KT}$.

To further characterize the properties above $T_{KT}$, we have studied with 2D XY model simulations the net winding number $W$ of $N^+$ vortices and $N^-$ antivortices that are thermally excited in a circular area of radius $R$, given by $W = N^+ - N^-$.  We find that the average squared winding number 
$\left\langle {{W^2}} \right\rangle $ increases linearly with $R$ at all temperatures, a perimeter law, in disagreement with an initial speculation of KT \cite{ktrev} (and taken up by others \cite{volovik}) that above $T_{KT}$ it should vary as $R^2$ (an area law).  However, our result is in complete agreement with 2D Coulomb-gas theories \cite{van,gruber,maryal,martin,Levesque}, which show that a perimeter law is expected, given that the  long-range interaction between vortex pairs is the same as the 2D Coulomb interaction.  We find that slope of the squared winding number with $R$ has a sharp peak near 1.15 $T_{KT}$, very similar to the specific heat peak.  The slope then decreases at $T = \infty$ to a value equal to an early theoretical prediction of Dhar \cite{dhar}.  

\begin{figure}[t]
\includegraphics[width=0.5\textwidth]{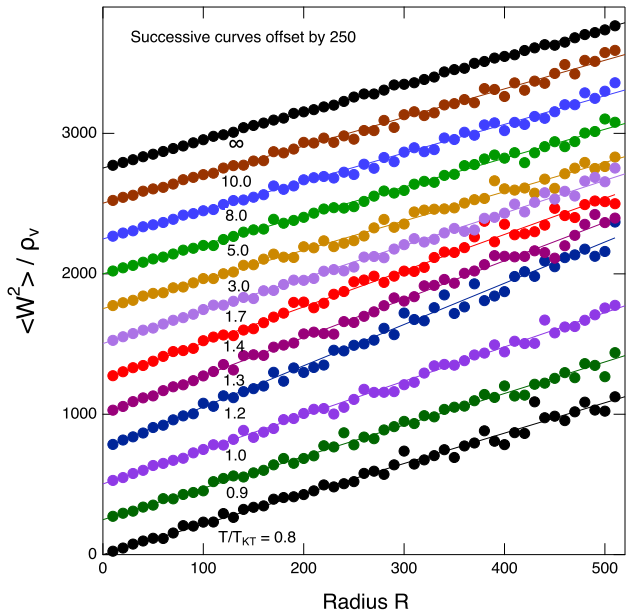} 
\caption{Average squared winding number divided by the vortex density $\rho_v$ in a circle of radius $R$.  The curves at different temperatures are offset by 250 on the vertical axis, otherwise they would nearly coincide.  Straight lines are linear fits to the data.}
\label{fig1}
\end{figure}

We have also studied above $T_{KT}$ the two-point vortex-vortex distribution functions \cite{ahns,still}, $\Gamma^{+-}$ and $\Gamma^{++} = \Gamma^{--}$, where $\Gamma^{+-}$ is the density of vortex-antivortex pairs per unit area of separation between $r$ and $r + dr$, and  $\Gamma^{++}$ the same for like-circulation pairs.  These are found to be asymptotic power-law decays at all temperatures, varying as $r^{x}$, and we find that the decay exponent for $\Gamma^{+-}$ has a peak  at a value $x \approx -3.2$ near 1.1 $T_{KT}$, close to the specific heat and winding-number peaks, while the $\Gamma^{++}$ exponent is $x = -4$ at all temperatures.  Above $T_{KT}$ these distribution functions are not the same as the vortex-vortex correlation functions, which are known to decay exponentially with $r$ \cite{minnhagen,samaj}.  Below $T_{KT}$, however, they have been shown to be the same as the correlation functions \cite{Frohlich,alacornu,alacornu2}, both having power-law decays.  This allows us to use a Coulomb-gas sum rule \cite{martin} with the distribution functions to verify the increase in the winding number slope below $T_{KT}$. 

The simulations of the 2D XY model use the standard Metropolis algorithm on a 1024$\times$1024 planar spin lattice with periodic boundary conditions.  The system is started from random spin orientations, and allowed to thermalize for 10$^6$ Monte Carlo steps.  To average the results 1000 configurations spaced by 50,000 Monte Carlo steps are recorded.  For each of those the vortex positions are identified by searching for net $\pm 2\pi$ rotations about neighboring spins, a well-known procedure from previous simulations \cite{binder,cugliandolo}.  The resulting averaged vortex density $\rho_v$ per lattice site agrees with those previous results, and approaches 1/3 per lattice site at infinite temperature (which is purely random spin configurations).  

From the vortex position maps the net winding numbers in circles from the lattice center are computed for radii (in lattice constants) $R$ = 10, 20...up to 500.  Squaring and dividing by the vortex density $\rho_v$, the results are shown in Figure \,\ref{fig1} for a range of temperatures between 0.8 $T_{KT}$ and infinity.  The successive curves are offset by 250 in the vertical axis since otherwise they would nearly coincide.  It is clear from the figure that the squared winding number is linear at all temperatures, a perimeter law.  KT  \cite{ktrev} speculated this would occur only below $T_{KT}$ where bound pairs near the perimeter would be ``cut" by the circular radius, contributing to the winding number as a perimeter law, while above $T_{KT}$ there would be a ``free" randomly-positioned vortex plasma state, where the squared winding number would be proportional to the total number of vortices within the circle, an $R^2$ area law.  However, it was later realized by Coulomb-gas theorists that this argument is not correct \cite{van,gruber,maryal,martin}, that actually any system with long-range forces between interacting objects will show a perimeter law for the fluctuations, and that this will hold at any temperature.  Our results show that this idea is correct for XY vortices, which interact with the long-range ln$(r/a_0)$ potential between a vortex and an antivortex separated a distance $r$, which is the same as the 2D Coulomb potential.  By extension, such a perimeter law should also hold in 2D superfluids where the vortex interactions are the same.

\begin{figure}[t]
\includegraphics[width=0.5\textwidth]{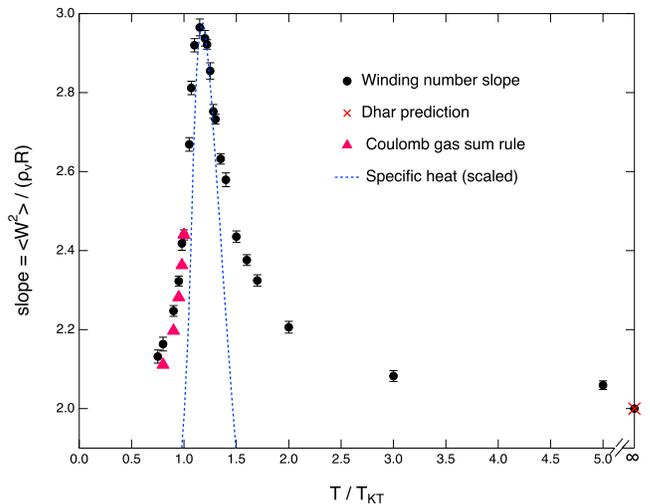} 
\caption{Slope of the squared winding number versus temperature, which shows a sharp peak near 1.15 $T_{KT}$, similar to the specific heat peak (blue dotted curve from Ref.\,\cite{pre_sh}, scaled to match the slope peak amplitude).  The Dhar prediction of Ref.\,\cite{dhar} is shown as the red cross, and the Coulomb-gas sum rule results (discussed in the text) are the red triangles.}
\label{fig2}  
\end{figure}

There is a noticeable change in the slope of the winding-number plot of Fig.\,\ref{fig1} near 1.2 $T_{KT}$.  We investigated this with more closely-spaced temperature intervals, and the resulting linear slopes are shown in Figure \ref{fig2}.  There is a sharp peak in the slope there, with a maximum near 1.15 $T_{KT}$.  This peak is quite similar to the specific heat peak \cite{pre_sh}, shown as the dotted blue curve in Fig.\,\ref{fig2}, which has been scaled by a factor to match the peak amplitude of the slope.  It is perhaps not surprising that these quantities are related, since the specific heat arises from energy fluctuations, while the winding number comes from vortex-number fluctuations.

Well above $T_{KT}$ the slope decreases to a value of 1.99$\pm$0.01 at $T = \infty$.  This is almost exactly equal to the value of 2.0 found in an early theory of vortex fluctuations by D. Dhar \cite{dhar}, who first predicted the perimeter law.  Since his work employed a square area, to compare with our circular area his slope of $\pi/2$ needs to be multiplied by $8/2\pi$, the ratio of the perimeter of a square of sides 2R to the perimeter of a circle of radius R.   As Dhar remarks, the existence of the perimeter law shows that the vortices are still strongly correlated even at infinite temperature, and are not randomly located.  In the XY model it is only the spin orientations that are completely randomized there, and correspondingly in a 2D superfluid the phase angle of the macroscopic wavefunction will be random, but not the vortices, which are formed from $\pm 2\pi$ rotations of the phase.

We have verified that the fluctuations of the winding numbers about $W = 0$ form a Gaussian distribution, and that the variance of the distribution satisfies a perimeter law \cite{maryal,Levesque}.  The data points in Fig.\,3 show the standard deviation of the Gaussian fits at $R$ = 500 as a function of temperature.  The solid line shows the prediction of Martin and Yalcin \cite{maryal}, $\sigma = (\left\langle {{W^2}} \right\rangle)^{1/2}$, where the values of $\left\langle {{W^2}} \right\rangle$ are taken from Fig.\,1 at $R$ = 500.  Since the squared winding number is proportional to $R$, the variance $\sigma^2$ will also increase with $R$, and we have verified that variation in Gaussian fits for different $R$.

The vortex-vortex distribution functions are computed following the methods outlined in Refs.\,\cite{still,
cugliandolo}.  Pairs of vortices with the shortest separations are tabulated and removed from the position map.  The next shortest pairs are similarly tabulated and removed, and this continues until all of the pairs are counted.  The pairs are sorted and counted into one lattice distance bins, giving the number of pairs $N(r)$ of separation between $r$ and $r+dr$.  The distribution function is then $\Gamma (r) = N(r)/2\pi r$. 
\begin{figure}[t]
\includegraphics[width=0.5\textwidth]{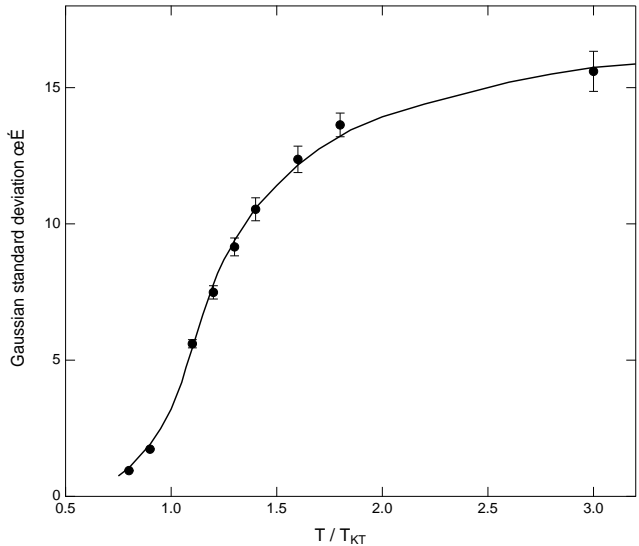} 
\caption{Standard deviation of Gaussian fits to the winding number distributions at different temperatures.  The solid line is the prediction of Martin and Yalcin \cite{maryal}.  }
\label{fig3}
\end{figure}

\begin{figure}[t]
\includegraphics[width=0.5\textwidth]{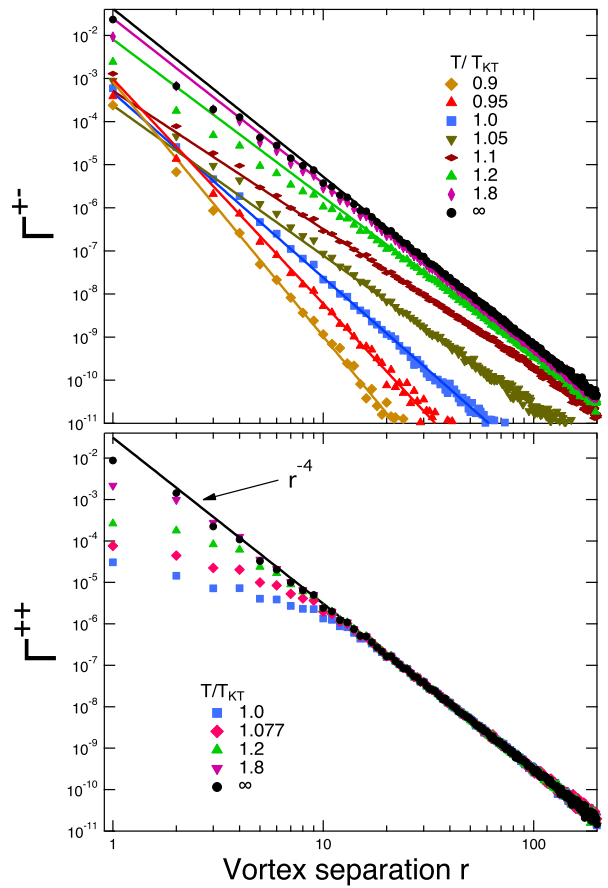} 
\caption{Vortex-vortex distribution functions versus their separation $r$, for  number of different values of $T/T_{KT}$.  The solid lines are asymptotic fits.}
\label{fig4}
\end{figure}

\begin{figure}[t]
\includegraphics[width=0.5\textwidth]{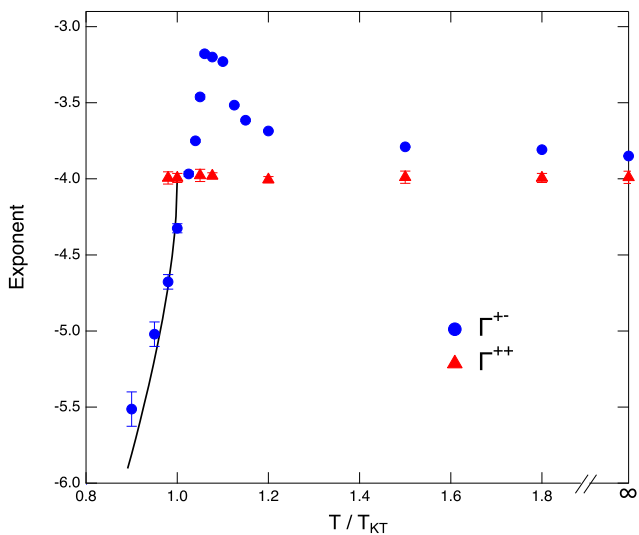} 
\caption{Exponents of the distribution functions versus temperature.  The solid line is the prediction of the KT theory \cite{kosterlitz}.}
\label{fig5}
\end{figure}

Figure \ref{fig4} shows the results for both   
$\Gamma^{+-}$ and $\Gamma^{++}$ for a range of temperatures.  The solid lines show the asymptotic power-law fits to the data, computed with the Python powerlaw package \cite{alstott}.  Figure \ref{fig5} shows the exponents from those fits versus temperature.  $\Gamma^{++}$ falls off asymptotically with an exponent $x = -4$ at all temperatures, while at short separations it is reduced due to the repulsive interactions between like-sign vortices, with the reduction being more pronounced at lower temperatures.  Below $T_{KT}$, $\Gamma^{+-}$ is known to fall off as $r^{-2\pi K}$ \cite{kosterlitz}, where for a superfluid $
K = {\hbar ^2}{\sigma _s}(T)/{m^2}{k_B}T$ with $\sigma _s$ the areal superfluid density, and $m$ the atomic mass.  The equivalent value of $K$ for the XY model is the helicity modulus \cite{binder}, and the resulting distribution prediction is shown as the solid line in 
Fig.\,\ref{fig5}, in general agreement with the data (the low vortex density in this region leads to large error bars in the distributions, and we neglect finite-size effects near $T_{KT}$).  Above $T_{KT}$, the exponent of $\Gamma^{+-}$ continues to increase, reaching a peak near 1.1 $T_{KT}$, similar to the slope and specific heat peaks, and then decreases at higher $T$ towards a value of -4 as $T \rightarrow \infty$.  We do find however, that the exponent of  $\Gamma^{+-}$ does not precisely go to -4 as $\Gamma^{++}$ does, but has a value of $-3.85\pm 0.01$ at $T = \infty$.  We speculate that this difference from -4.0 could be a finite-size effect.  Coulomb-gas theories have predicted that finite-size effects will be present at all temperatures for long-range systems \cite{forrester}, instead of only very close to a phase transition as found in more usual thermodynamic systems.  We do not understand, however, why this seems not to be the case for $\Gamma^{++}$. 

These results for the distribution functions further illustrate that the vortex positions are not random at high temperatures.  We have checked that randomly positioning 2D vortices and antivortices leads to an asymptotic exponent of -3.0.  This is clearly not the case for $\Gamma^{+-}$, except possibly near the ``unbinding" peak at 1.1 $T_{KT}$, where the exponent seems to approach -3.  

We are able to establish a connection between the distribution functions and the peak in the winding number slope below $T_{KT}$ using a Coulomb-gas sum rule \cite{van,gruber,maryal,martin}.  In this region the vortex-vortex correlation functions follow the power-law decays seen in Figs.\,\ref{fig4} and \ref{fig5}, and have been shown to be identical to the distribution functions \cite{Frohlich,alacornu,alacornu2}.  Thus we can insert 
$\Gamma^{+-}$ into the sum rule, and it can then be written as
\begin{equation}
\frac{{\left\langle {{W^2}} \right\rangle }}{{2\pi R}} = \frac{2}{\pi }\int {r\,{\Gamma ^{ +  - }}\,} 2\pi r\,dr \quad.
\end{equation}
By numerically integrating this equation and dividing by the vortex density, the resulting values for the winding-number slope are shown as the red triangles in Fig.\,\ref{fig2}.  It can be seen that this successfully models the initial increase in the slope in the region below $T_{KT}$.  Above $T_{KT}$, however the sum rule fails completely, giving slopes too large and that stay relatively constant in temperature, probably reflecting the change in the vortex-vortex correlation functions to exponential decay \cite{minnhagen,samaj}.  With the peak in the exponent of 
$\Gamma^{+-}$ coming close to the peak in the slope it would seem likely that there would be some type of sum rule for the distribution functions that would model the slope at all temperatures, but this remains unknown.  We are unaware of any theories that might connect the correlation and distribution functions above $T_{KT}$.

Our results have major implications for the dynamics of the decay of vorticity in temperature-quenched 2D superfluids.  For instantaneous quenches starting from initial temperatures below $T_{KT}$ it has been shown that the initial separation dependence of $\Gamma^{+-}$ plays a fundamental role in the dynamics following the quench \cite{forrester2013}.  At lower starting temperatures the more negative exponent of $r^{x}$ in $\Gamma^{+-}$ gives rise to a more rapid decay of the initial vortex density, varying with time $t$ after the quench as $t^{(x+2)/2}$.  The interpretation of this is quite simple:  a more rapid decrease of $\Gamma^{+-}$ means most of the vortex pairs have relatively small separations, and hence diffuse together and annihilate more rapidly.  If this continues to hold for instantaneous quenches from above $T_{KT}$, as we might expect, the exponents shown in Fig.\,\ref{fig5} mean that the time decays will be slower than $t^{-1}$, and will vary considerably depending on the initial temperature.  Indeed the temperature quench in XY simulations \cite{cugliandolo} starting from 2.0 $T_{KT}$ found a variation near $t^{-0.7}$.  It would be interesting to see if a similar quench from 1.1 $T_{KT}$ would give a time decay exponent closer to -0.5, and then a quench from $T = \infty$ (with $x \approx -4$) would give something close to -1.0, the value expected from dynamic scaling \cite{forrester2013,bray}.  

In summary, we have studied fluctuations of thermally excited vortices above and below the Kosterlitz-Thouless superfluid transition, finding a perimeter law for winding number fluctuations, and have characterized new peaks found in the winding number and in the vortex-antivortex distribution functions near the specific heat peak.  The results are in full agreement with Coulomb-gas theories, and show that strong correlations between the vortices still exist even at $T = \infty$, as first postulated by Dhar.

\begin{acknowledgments}
We thank Prof.\,Dhar for pointing out the numerical factor between his theory and our results.  We also thank the Hummingbird Team at UCSC, especially Rion Parsons and Josh Sonstrom, for their support and for the use of the Hummingbird Cluster in this research.  This work was supported in part by a grant from the Julian Schwinger Foundation.
\end{acknowledgments}

\section*{References}

\end{document}